\UseRawInputEncoding
\documentclass[journal=jacsat,manuscript=article]{achemso}

\usepackage{hyperref}

\usepackage[version=3]{mhchem} 

\author{Vandana Sharma}
\email{vandana.sharma@students.iiserpune.ac.in}
\affiliation{Department of Physics, Indian Institute of Science Education and Research, Pune-411008, India}
\author{Sunny Tiwari}
\affiliation{Department of Physics, Indian Institute of Science Education and Research, Pune-411008, India}
\author{Diptabrata Paul}
\affiliation{Department of Physics, Indian Institute of Science Education and Research, Pune-411008, India}
\author{Ratimanasee Sahu}
\affiliation{Department of Physics, Indian Institute of Science Education and Research, Pune-411008, India}
\author{Vijayakumar Chikkadi}
\affiliation{Department of Physics, Indian Institute of Science Education and Research, Pune-411008, India}
\author{G.V. Pavan Kumar}
\email{pavan@iiserpune.ac.in}
\affiliation{Department of Physics, Indian Institute of Science Education and Research, Pune-411008, India}

\title{Optothermal pulling, trapping, and  assembly of colloids using nanowire plasmons}

\begin{document}


%

\begin{abstract}
Optical excitation of colloids can be harnessed to realize soft matter systems that are out of equilibrium. In this paper, we present our experimental studies on the dynamics of silica colloids in the vicinity of a silver nanowire propagating surface plasmon polaritons (SPPs). Due to the optothermal interaction, the colloids are directionally pulled towards the excitation point of the nanowire. Having reached this point, they are spatio-temporally trapped around the excitation location. By increasing the concentration of colloids in the system, we observe multi-particle assembly around the nanowire. This process is thermophoretically driven and assisted by SPPs. Furthermore, we find such an assembly to be sensitive to the excitation polarization at input of the nanowire. Numerically-simulated temperature distribution around an illuminated nanowire corroborates sensitivity to the excitation polarization. Our study will find relevance in exploration of SPPs-assisted optothermal pulling, trapping and assembly of colloids, and can serve as test-beds of plasmon-driven active matter.
\end{abstract}


\section{Introduction}
How do colloids\cite{palacci2013living}\cite{lozano2016phototaxis}\cite{buttinoni2012active}\cite{doi:10.1126/science.1253751}behave in the vicinity of an optical and/or optothermal potential? How can colloidal dynamics and assembly be controlled by harnessing optical excitation? These are relevant questions in driving soft matter out of equilibrium, where the driving force is provided by an optical excitation. In this study, we present directional pulling of dielectric colloids in the vicinity of an optically-excited metal nanowire. This movement further leads to spatio-temporal trapping of the same colloids. Upon increasing the concentration of the colloids, a two dimensional assembly emerges. This emergent process is found to be sensitive to the excitation polarization of the light illuminating the nanowire.

In this context, understanding the dynamics of colloids under such optical excitation schemes is relevant.\cite{zaidouny2013light}\cite{PhysRevLett.81.2606}\cite{schmidt2019light} \cite{jenkins2008colloidal}\cite{dobnikar2013emergent}Recently, colloidal manipulation using optical pulling forces has also gained significant interest.\cite{chen2011optical}\cite{lu2017light}\cite{ali2020tailoring}\cite{li2020optical} Typically, colloids undergo diffusive Brownian motion. \cite{bian2016111} Brownian motion is relevant to the understanding of the dynamics of soft matter and biological systems \cite{frey2005brownian} as well as fabricated structures designed to mimic naturally occurring systems.\cite{lozano2016phototaxis}\cite{lozano2018run}\cite{niese2020apparent} The study of the physics of systems out of equilibrium \cite{blickle2007einstein}\cite{ganguly2013stochastic} is especially of pertinent interest as most of the living matter is far from equilibrium. In this context, colloids have been extensively studied to explore the dynamics of such systems.\cite{RevModPhys.88.045006}\cite{rings2010hot}\cite{su2021towards}  \cite{fernandez2020feedback} The colloids have the advantage of easy manipulation under external fields. \cite{caciagli2020controlled}\cite{vialetto2021photothermally} \cite{arya2021light}\cite{ghosh2019all} \cite{tkachenko2020light}\cite{ghosh2019self}\cite{ghosh2020directed}Conventionally, colloids are trapped using a focused laser beam\cite{ashkin1986observation}\cite{jones2015optical}. Although powerful, the conventional optical schemes are constrained by the diffraction limit of light.\cite{melzer2018fundamental} As a complement to this, surface plasmon polaritons (SPPs) in metallic nanostructures can facilitate radiative and non-radiative pathways below the diffraction limit of light. \cite{sharma2020large}\cite{patra2014plasmofluidic}\cite{patra2016large}\cite{garces2006extended} SPPs are surface electromagnetic waves at metal-dielectric interface.\cite{barnes2003surface} They decay into the dielectric medium through radiative and non-radiative channels. Radiative channels can be harnessed for optical trapping and assembly through SPPs momentum, and non-radiative channels of SPPs can be used for optothermal assembly via thermophoretic\cite{piazza2008thermophoresis}\cite{piazza208thermophoresis}\cite{reichl2014charged} interactions.

Of relevance to this study is SPPs propagation along the silver nanowire.\cite{shegai2011unidirectional} \cite{yang2016identification} When illuminated with a focused laser beam, these single crystalline, chemically-prepared nanowires can propagate SPPs. Given the geometry of the nanowire, quasi-one dimensional confinement of plasmons can lead to interesting optical and optothermal effects at sub-wavelength scales. Although SPPs in the silver nanowire have been studied extensively in the context of nanophotonics\cite{guo2019routing}\cite{johns2017dynamics}\cite{vasista2018differential}\cite{tiwari2020dielectric}\cite{doi:10.1021/acs.jpclett.1c01923}, their interface with soft-matter systems such as colloids is yet to be explored in detail. \cite{yang2016guided} \cite{nan2019silver}  SPPs have been utilized in the past to assemble particles on a large scale. But such studies have mainly relied on extended 2D plasmonic platforms such as metallic films.\cite{patra2014plasmofluidic}\cite{garces2006extended} 

Motivated by this, herein we report the directed pulling and trapping of a single silica colloid as well as assembly formation of colloids by utilizing a single, silver nanowire to create a non-equilibrium environment in the system. A laser beam (532 nm) is focused at the end of a silver nanowire, which leads to the excitation of SPPs. In this study, we explore the transport and trapping of single colloids (figure 1a), and multiparticle assembly (figure 1b) due to the optical excitation of the plasmonic silver nanowire. A representative optical image of SPPs propagating in a silver nanowire at 532 nm  is shown in figure 1c(i). The excitation polarization of the focused laser beam is aligned along the nanowire axis. This leads to SPPs propagation which out-couples at the distal end of the nanowire (dotted circle in figure 1c(i)). The SPPs supported by the nanowire generate heat\cite{baffou_2017} \cite{baffou2013thermo}\cite{moller2018fast}which is subsequently released into the surrounding environment creating a temperature gradient in the system. The subsequent dynamics of the particles is studied by looking into the mean square displacement (MSD) of the particles. The MSD at time $\tau$ is defined as the ensemble average: 
\begin{equation} 
MSD (\tau) = <\Delta r(\tau)^2> = <[r(t+\tau) - r(t)]^2>.
\label{equation}
\end{equation}
where, $r(t)$ is the position of the particle at time t, $\tau$ is the time lag between two positions of the particle whose displacement is $\Delta r(\tau),<>$ is the time-average over t and/or an ensemble-average over several trajectories.

\begin{figure}[H]
\centering
  \includegraphics[width=\linewidth]{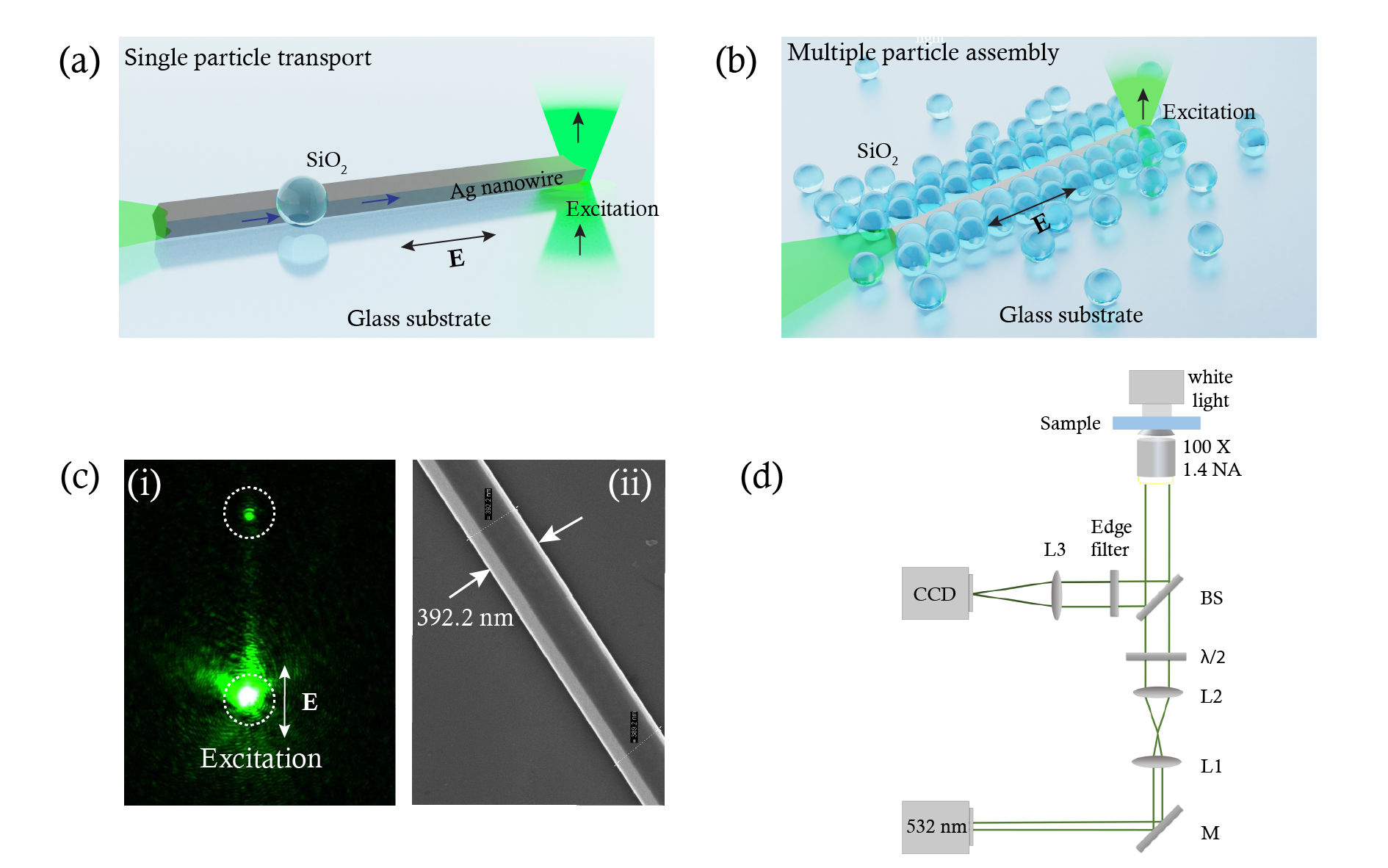}
  \caption{Schematic illustration of the experimental  configuration. Silver nanowires are dropcasted on the glass substrate. One end of the Ag nanowire is excited using a 532 nm laser to excite the SPPs. The light out-couples from the distal end of the wire. Silica colloids dispersed in milli-Q water are dropcast on the glass substrate. The whole system is enclosed in a 120  $\mu$m chamber. (a) Schematic illustration of the single silica particle transport along the silver nanowire.  (b) Schematic illustration of the multiple particle assembly along the silver nanowire. (c) (i) Excitation of SPPs along the silver nanowire when the polarization of the input laser is along the length of the nanowire.  (c)(ii) SEM image of a representative silver nanowire used in the experiments. (d) Schematic representation of the experimental setup. The silver nanowire is excited using a high numerical aperture 100 X, 1.4 NA objective lens. M is the mirror. The laser beam is expanded using a combination of lenses, L1 and L2. The polarization of the input laser is controlled using a half wave plate. The sample is illuminated from the top using a white light. The video is recorded by CCD after rejecting the elastically scattered light using a 532 nm edge filter.}
  
\end{figure}

\section{Methods}

\subsection{Sample preparation and experimental design}

Our experimental design constitutes a  simple geometry wherein a silver nanowire is dropcasted on a glass substrate and left to dry. The silver nanowires are chemically synthesized using the polyol process. \cite{sun2003polyol} The typical size of the silver nanowire used in the experiments is $\approx$ 300-400 nm in diameter and $\approx$12 $\mu$m in length. The nanowire is characterized using scanning electron microscopy as shown in figure 1(c)(ii). The colloids consist of 2 $\mu$m silica beads (purchased from Microspheres-Nanospheres) and 2.2 $\mu$m polystyrene beads (purchased from micro particles Gmbh).  The colloids dispersed in milli-Q water are dropcasted on the glass substrate. The whole system is enclosed in a 120 $\mu$m spacer.  Figure 1 (d) shows the schematic of the inverted experimental setup. A 532 nm laser is expanded using a combination of two lenses L1 and L2. A half wave plate is used to control the input polarization of the beam. One end of the silver nanowire is excited by tightly focusing the laser using  a 100 X, 1.4 NA objective lens. The power at the sample plane is measured to be 5 mW. The polarization of the incoming laser is kept along the nanowire unless specified otherwise, for efficient excitation of the SPPs. The videos are recorded by the CCD at 20 fps after rejecting the elastically scattered light by a 532 nm edge filter.

\subsection{Numerical simulations}
The temperature distribution of the system is calculated using Finite element method based numerical simulations (COMSOL Multiphysics) by solving electromagnetic wave and heat transfer in solids and fluids modules in 3D. An 8 $\mu$m long silver nanowire with a diameter of 350 nm is excited using focused 532 nm gaussian beam. The whole geometry is simulated using free tetrahedral user defined mesh.

\section{Results and Discussion}

\subsection{Single particle transport}

A single nanowire dropcasted on a glass substrate can act as a guiding medium for the directed motion of the colloid from its captured location to the excitation point. The nanowire initially traps the particle at one of the locations of the wire depending on the initial position of the particle with respect to the wire. 

Here, we demonstrate the trapping of the silica colloid from the distal end of the nanowire. If the particle is closer to any other location of the wire when it is in its vicinity, it will get trapped there first, and then move towards the excitation point. One end of the silver nanowire is excited through a 100 X, 1.4 NA objective lens. The white dotted circle represents the excitation point. Figure 2(a) shows the time series images of a silica colloid transport along the nanowire. At t= 0 s , the colloid is freely diffusing in the water. As it comes closer to the wire, it gets trapped at the distal end at 13 s and starts to move along the nanowire. Within 65 s, the particle is trapped at the excitation point (See supplementary movie S1). To analyse the dynamics of the colloidal particle, it is essential to map out the trajectory and the associated MSD of the particle. 

\begin{figure}[H]
\centering
  \includegraphics[width=1.0\linewidth]{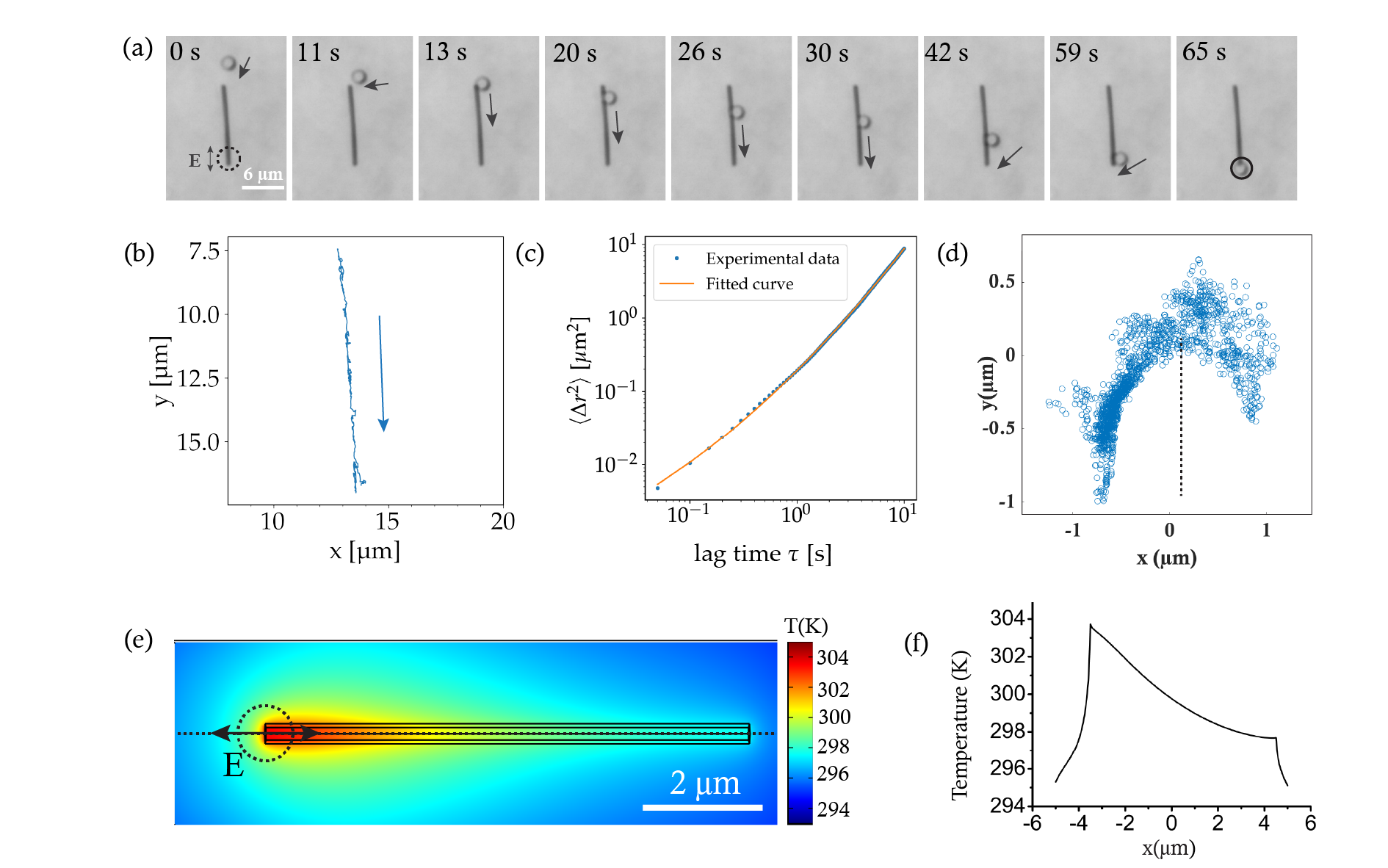}
  \caption{Single particle transport. (a) Time series images of the directed motion of the silica  colloid along the silver nanowire from the distal end towards the excitation point of the laser. The polarization of the input laser is kept along the length of the silver nanowire for efficient excitation of SPPs. Black arrows indicate the direction of movement of the particle. The particle gets transported from one end of the nanowire to the other end within a minute. (b) The trajectory of the colloid is plotted. The blue arrow indicates the direction of particle transport. (c) Mean square displacement of the colloid. The blue dotted curve is the experimental data and orange curve is the fit. (d) Trajectory of the particle once it is confined at the excitation point. The dotted line represents the orientation of the nanowire with respect to the particle trajectory. (e) Temperature distribution of the system in the x-y plane at z=0 i.e. at the base of the nanowire. (f) A crosscut along the top of the nanowire at z=317 nm shown in (e) as the black dotted line. The 0 in the graph corresponds to the centre of the nanowire.}
  
\end{figure}

\pagebreak

Since the particle continuously changes its micro-environment from diffusive to moving along the nanowire to permanent trapping at the excitation point, it is crucial to analyze the sections separately and not mix the resulting dynamics.  Consequently,the particle is tracked from the moment it makes contact with the wire till it is transported to the other end before it gets trapped at the excitation location using Trackpy.\cite{allan_daniel_b_2021_4682814} The associated trajectory of the particle is plotted in figure 2(b). The trajectory plot shows that the motion of the particle is highly directional. The blue arrow indicates the direction of the movement of the particle. To further investigate the nature of the colloidal transport, we plot the MSD of the colloid on a log-log scale for the same time interval, as shown in figure 2(c). The experimentally obtained data is plotted in dotted blue. The motion of the colloid on the nanowire can be considered as a combination of 1D Brownian motion and ballistic motion. The MSD for such a system can be described as\cite{howse2007self}: 
\begin{equation} 
MSD (\tau) = <\Delta r(\tau)^2> = v^2 \tau^2 + 2D\tau.
\label{equation}
\end{equation}
where, \textit{v} is the terminal  velocity of the colloid due to the external force which pulls it towards the nanowire and D is the diffusion coefficient. An analysis of the velocity of silica colloid while it is being transported on nanowire over time (SI figure 1) shows that its velocity fluctuates roughly around a constant value. By fitting equation 2 to the experimentally observed data in figure 2(c), we get \textit{v} as 0.27$\mu$ms$^{-1}$ and a diffusion coefficient of 0.07 $\mu$m$^2$s$^{-1}$.  A comparative study for the Brownian motion of the ensemble of colloids which are under no laser excitation is given in SI figure 1(see supplementary movie S2). Other examples of particle-wire transport system are given in SI figure 3.

A previous study reported the transport of a much smaller TiO$_2$ particle along the nanowire, but it is not permanently trapped and is pushed away from the excitation point into the solution. \cite{yang2016guided} In our case, once the silica colloid reaches the excitation point, it is permanently trapped there as long as the excitation laser is on. The trajectory of the trapped particle is shown in figure 2(d). By turning off the laser, the particle can again be released into the solution.

The efficient excitation of SPPs is crucial for the transport of the colloid. When a metallic structure is irradiated by an electromagnetic field, there is an associated heat generated in the system due to Joule heating.\cite{baffou_2017} After plasmons are launched along the nanowire, heat is generated at the excitation point as well as along the length of the wire due to plasmon dissipation. Consequently, the temperature of the nanowire increases and the heat is released to the environment.  The temperature distribution details were obtained through COMSOL. The numerically calculated temperature distribution in the xy plane at z=0 is shown in figure 2 (d). The system attains a total temperature increment of $\approx$10 K at the excitation point. A plot of the temperature distribution at the top of the wire at z=317 nm is plotted in figure 2 (e) which shows that the temperature is highest at the excitation point of the wire and gradually decreases along the length of the nanowire.
As reported previously\cite{sharma2020large}, silica colloids show thermophoretic migration from a lower temperature to a higher temperature. When the silver nanowire is excited, a temperature gradient is set up in the nanowire as well as its surroundings. Initially, the particle is moving freely in the water. When the colloid comes into the field of optothermal gradient of the system, it senses the temperature gradient and and is pulled towards the excitation point having the highest temperature. 
 \newline
To compare with a particle of different composition, and to emphasize the role of thermophoretic interactions, experiments were also performed at the same laser power with the polystyrene (PS) beads of size 2.2 $\mu$m (See supplementary movie S3). PS beads have been previously reported to move from a higher temperature to a lower temperature.\cite{duhr2005two}\cite{braun2014trapping} In our experiments, the PS particles initially come closer to the nanowire but as they come too close, they are repelled from the excitation point. The selective trapping of the particles shows that thermophoresis has a contributing factor in trapping the particle.

\subsection{Confined dynamics of multiple colloids}  
A silver nanowire can not only trap and transport a single colloid but can also be used to assemble multiple particles along its length. A concentrated solution of silica colloids in milli-Q water was prepared so that multiple particles can come into the field of optothermal potential simultaneously. Figure 3 (a) shows the time series images of the gradual assembly of the colloids along the length of the nanowire. At 0 s, two silica particles are trapped at the excitation point. Within 32 s, ten particles are aligned along the length of the nanowire. 

\begin{figure}[H]
\centering
  \includegraphics[width=0.95\linewidth]{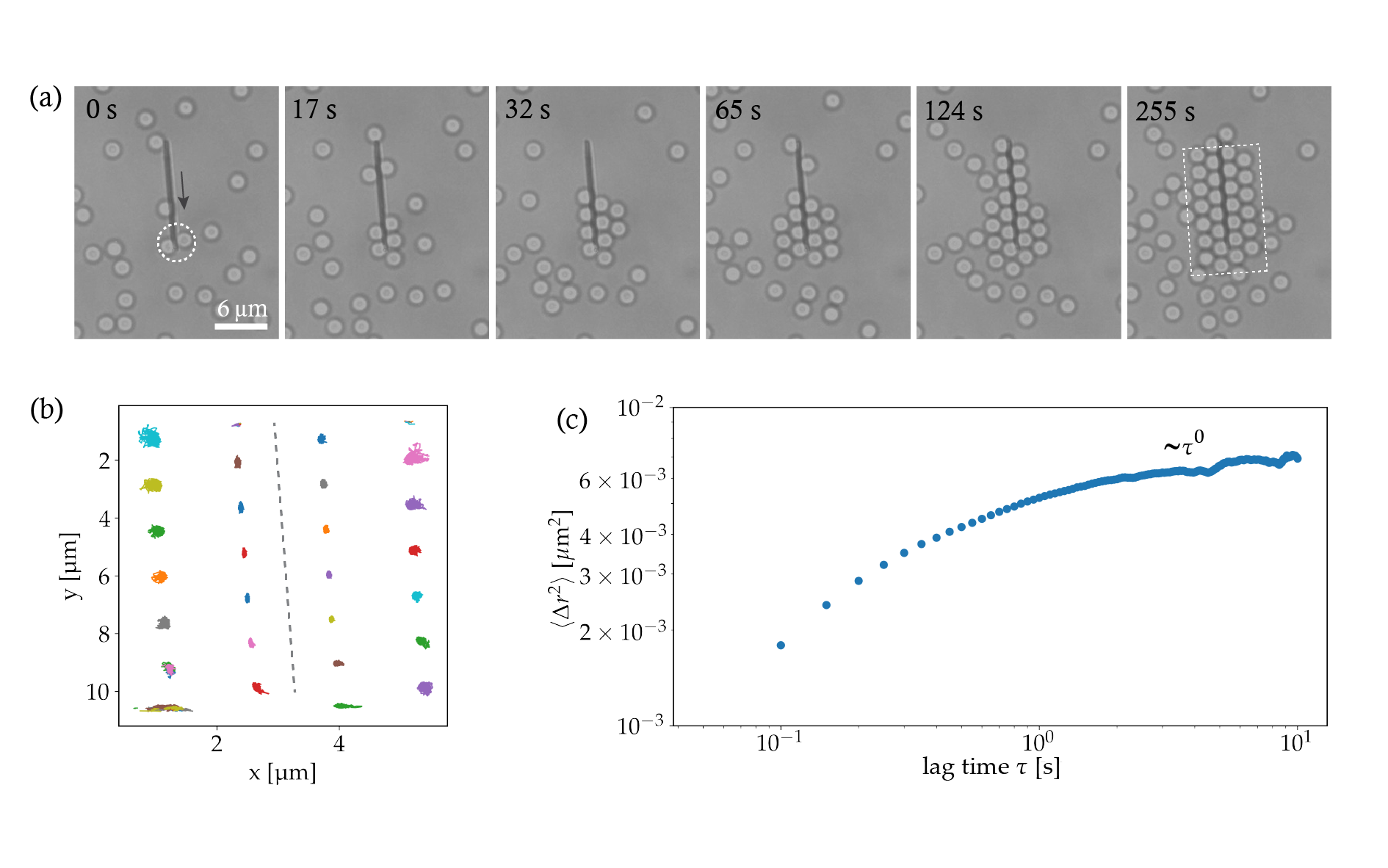}
  \caption{Confined motion of multiple particles. (a) Time series images of the gradual assembly of the colloids. Multiple layers of the colloids are formed along the length of the nanowire within 255 s. The dotted circle indicates the laser excitation point. The black arrow at 0 s indicates the direction of the motion of the colloids. The white dotted region indicates the region where colloids are tightly bound. (b) The trajectories of the particles in the white dotted region are plotted over a time period of 150 s. (c)  The averaged MSD of all the particles is plotted as a single curve.}
  
\end{figure}

As shown in figure 2 (e), the temperature is higher towards the first half of the wire closer to the excitation point. Consequently, the particles first accumulate near the excitation end of the nanowire and start forming the second layer on both sides of the wire at 65 s. At 255 s, we observe a two layer formation of the  colloids, which are stably trapped there as long as the laser excitation is on. The trajectories of the multiple particles  shown in dotted white box in figure 3(a) over a period of 150 s are plotted in figure 3 (b). See supplementary movie S4 for the entire assembly process. The first layer closer to the wire exhibits stronger confinement to the nanowire as can be seen through the smaller spread of the trajectories of the individual colloids. Moreover, the trajectories are more confined perpendicular to the wire compared to along the wire. Since, the second layer of the colloids are not bounded by a third layer, they exhibit a larger spread in the trajectories perpendicular to the wire. The averaged MSD of all the particles is plotted in figure 3 (c). The curve almost falls flat and saturates at large lag times with the  diffusion exponent of $\approx$ 0 indicating the confined motion of the trapped particles. The movement of the particles is impeded and they can not freely diffuse away from the trapped location.

\subsection{Polarization dependence of the assembly}
The heat generated by the silver nanowire can be contributed to two factors. One is absorption by the wire, and the other is due to the generation of SPPs\cite{li2019photothermal}. As mentioned previously, the SPPs play a crucial role in the generation of heat and consequently trapping the colloidal particles. Keeping the polarization of the beam along the silver nanowire leads to efficient excitation of SPPs as can be seen in figure 4 (a)(ii) and consequently, should lead to faster assembly process. Figure 4(a) and 4(b) compares the extent of the assembly for the two cases, one when the polarization of the incoming beam is along the length of the nanowire and second, when it is perpendicular to it. 

First the polarization is kept along the nanowire and the snapshot of the assembly is taken at 7 minutes as shown in figure 4(a)(i). Once the assembly reaches the saturation point, the laser is switched off and the colloids are free to redisperse into the solution. Now, the polarization of the laser is switched to perpendicular to the wire. In this case the efficiency of excitation of propagating plasmons is drastically reduced and no outcoupling of the light from the distal end is observed as can be seen in figure 4(b)(ii). A snapshot of the assembly after the same time of 7 minutes is shown in figure 4 (b)(i). A few particles near the excitation point and along the length of the wire can be seen but there is no significant assembly of the particles. The numerically calculated temperature distribution in the xy plane at z=0 is shown in figure 4(c). A comparative temperature distribution for the both the polarizations at the top of the wire is plotted in figure 4 (d). The assembly process for both the polarization can be found in supplementary movie S5 and movie S6.

\begin{figure}[H]
\centering
  \includegraphics[width=0.95\linewidth]{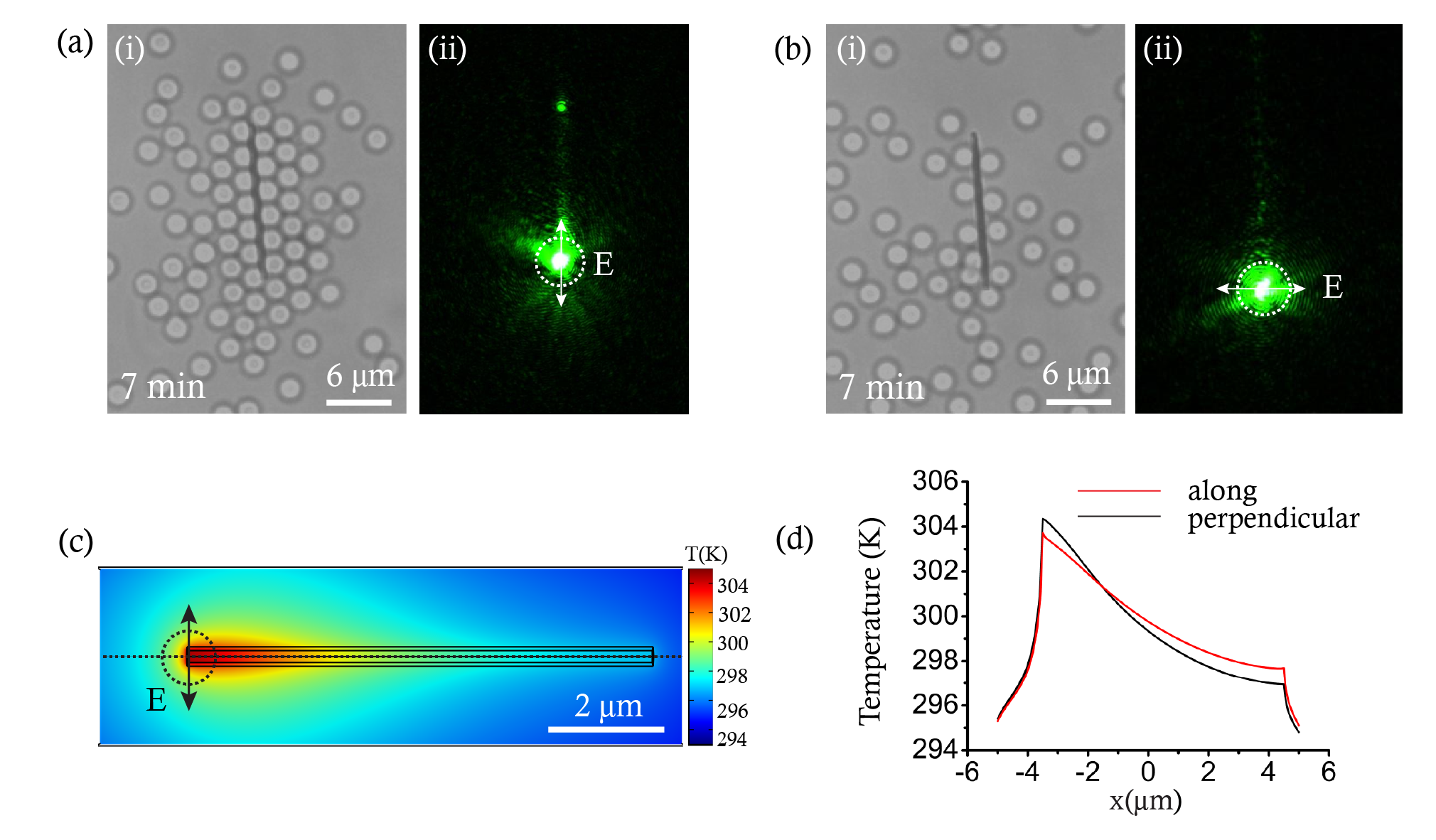}
  \caption{Polarization dependence of the assembly. (a) (i) Assembly of the colloids after 7 minutes when the polarization of the input laser is along the length of the silver nanowire. (ii) Excitation of silver nanowire. The light out-couples from the distal end of the wire. The dotted circle indicates the laser excitation point.  (b)(i) There is no significant assembly of the colloids though few particles are trapped near the excitation point. (ii) Excitation of the wire when the polarization is perpendicular to the long axis of the wire. (c) Temperature distribution in the x-y plane at z=0, i.e. at the base of the silver nanowire. (d) A crosscut of the temperature distribution along the black dotted line at z= 317 nm, i.e. at the top of the nanowire is plotted for both the orientations of the polarization. The 0 in the graph corresponds to the centre of the nanowire. }
  
\end{figure}

A  higher temperature increase at the excitation point is found in the case of perpendicular polarization. It can be attributed to the fact that at the excitation point the electric field is confined to a smaller region along the diameter of the wire leading to more localized plasmon excitation. The near-field electric field distribution for both the cases is shown in supplementary figure S3. This enhances the temperature increment at this point, but the gradual fall in the temperature is more  as there is less efficient excitation of plasmons along the nanowire and the resultant temperature increase due to plasmon dissipation is missing when there are no propagating plasmons.

\section{Conclusion}
We have experimentally studied the transport of the silica colloid by utilizing SPPs of a silver nanowire. Such a pulled colloid is spatio-temporally trapped at the nanowire excitation. Upon increasing the concentration of the colloids, we observed extended assemblies whose collective dynamics was studied and shown to be confined in nature. This assembly process is sensitive to the excitation polarization at nanowire input.This polarization-sensitivity was also corroborated in the simulated temperature distribution around the nanowire, thus indicating an intricate combination of optical and thermal effects facilitated by the nanowire SPPs. Given that the colloids are driven out of equilibrium by a quasi-one dimensional optothermal substrate, they may serve as interesting test-beds for active and driven matter at sub-micron scales. We envisage the utility of SPPs to create intriguing optothermal potentials to drive soft matter out of equilibrium, that can complement and improvise on conventional optical schemes.

\begin{acknowledgement}

Authors thank Pragya Kushwaha, Shailendra Kumar Chaubey and Chetna Taneja for fruitful discussions.  This work was partially funded by Air Force Research Laboratory grant (FA2386-18-1-4118 R\& D18IOA118) and Swarnajayanti fellowship grant (DST/SJF/PSA02/2017-18) to G V PK.

\end{acknowledgement}

\begin{suppinfo}

The videos can be found here 

\url{https://www.youtube.com/playlist?list=PLVIRTkGrtbrszZdisE9AewKpvJABfp02f}

\end{suppinfo}

\bibliography{achemso-demo}

\pagebreak

\begin{center}
  \textbf{\large Supplementary Material\\Optothermal pulling, trapping, and  assembly of colloids using nanowire plasmons}\\[.2cm]
  Vandana Sharma,$^{*}$ Sunny Tiwari, Diptabrata Paul, Ratimanasee Sahu, Vijayakumar Chikkadi and G. V. Pavan Kumar$^{*}$\\[.1cm]
  {\itshape Department of Physics, Indian Institute of Science Education and Research, Pune-411008, India}\\
  ${}^*$Email: vandana.sharma@students.iiserpune.ac.in, pavan@iiserpune.ac.in\\
\end{center}

\setcounter{equation}{0}
\setcounter{figure}{0}
\setcounter{table}{0}
\renewcommand{\theequation}{S\arabic{equation}}
\renewcommand{\thefigure}{S\arabic{figure}}
\renewcommand{\bibnumfmt}[1]{[S#1]}
\renewcommand{\citenumfont}[1]{S#1}

\pagebreak
\tableofcontents

\appendix
\renewcommand{\thesection}{S\arabic{section}}
\section{}
\renewcommand{\thefigure}{S\arabic{figure}}

\subsection{S1: Velocity of the colloid at different frame differences (df)}
\subsection{S2: Brownian motion of particles}
\subsection{S3: Single particle transport along the nanowire}
\subsection{S4: Near-field electric field distribution}
\subsection{S5: Details of supplementary movies}

\pagebreak
\section{S1: Velocity of the colloid at different frame differences (df)}

\begin{figure}[H]
\centering
  \includegraphics[width=\linewidth]{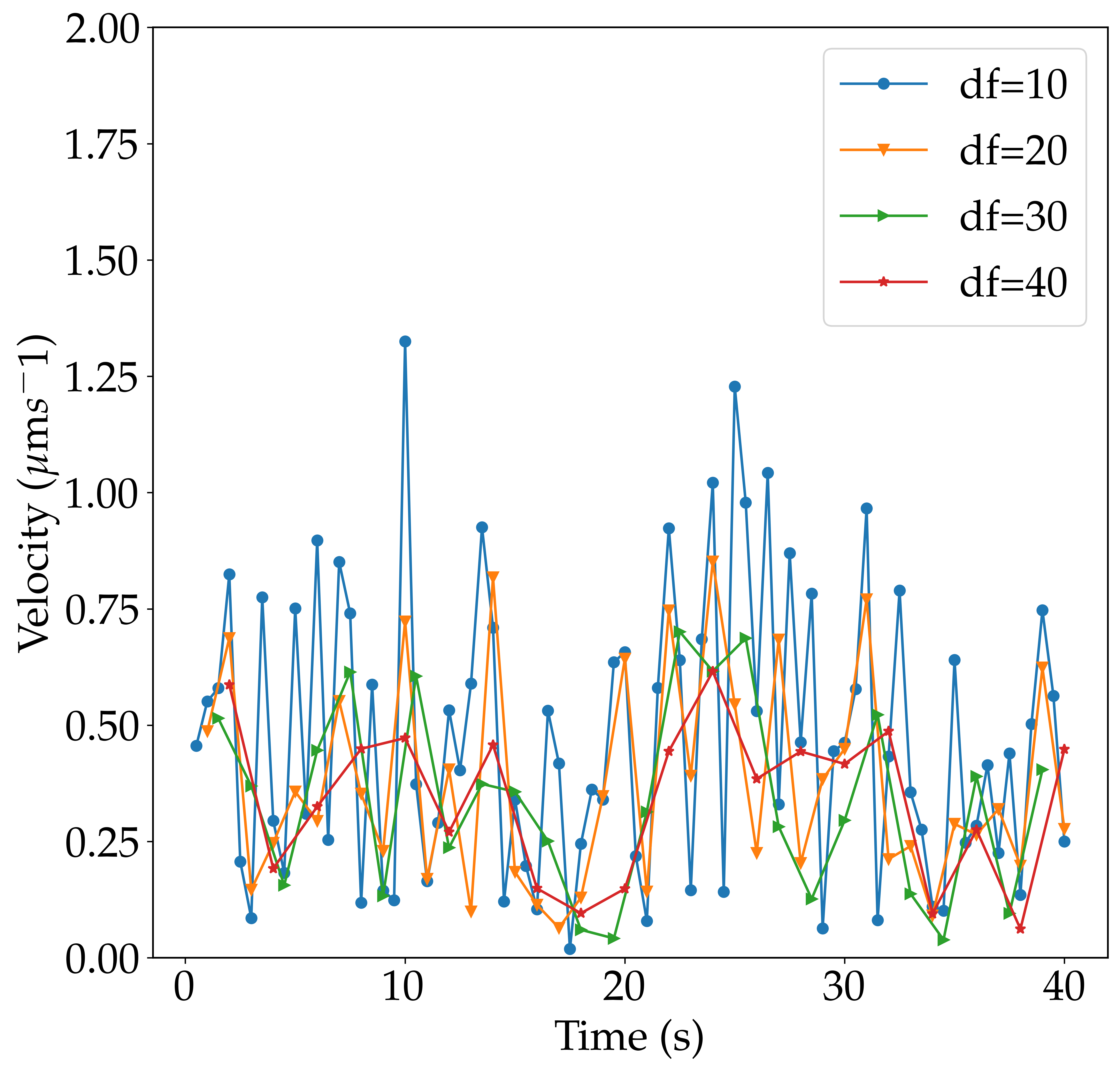}
  \caption{Displacements are calculated for frame differences (df) of 10, 20, 30, and 40 frames and corresponding velocities are plotted.}
  
\end{figure}

\section{S2: Brownian motion of particles}
2 $\mu$m Silica colloids in milli-Q water are dropcast on a glass substrate and sealed in a 120 $\mu$m spacer. The Brownian motion of the particles is recorded over a period of 50 s in the absence of the nanowire and laser excitation.

\begin{figure}[H]
\centering
  \includegraphics[width=\linewidth]{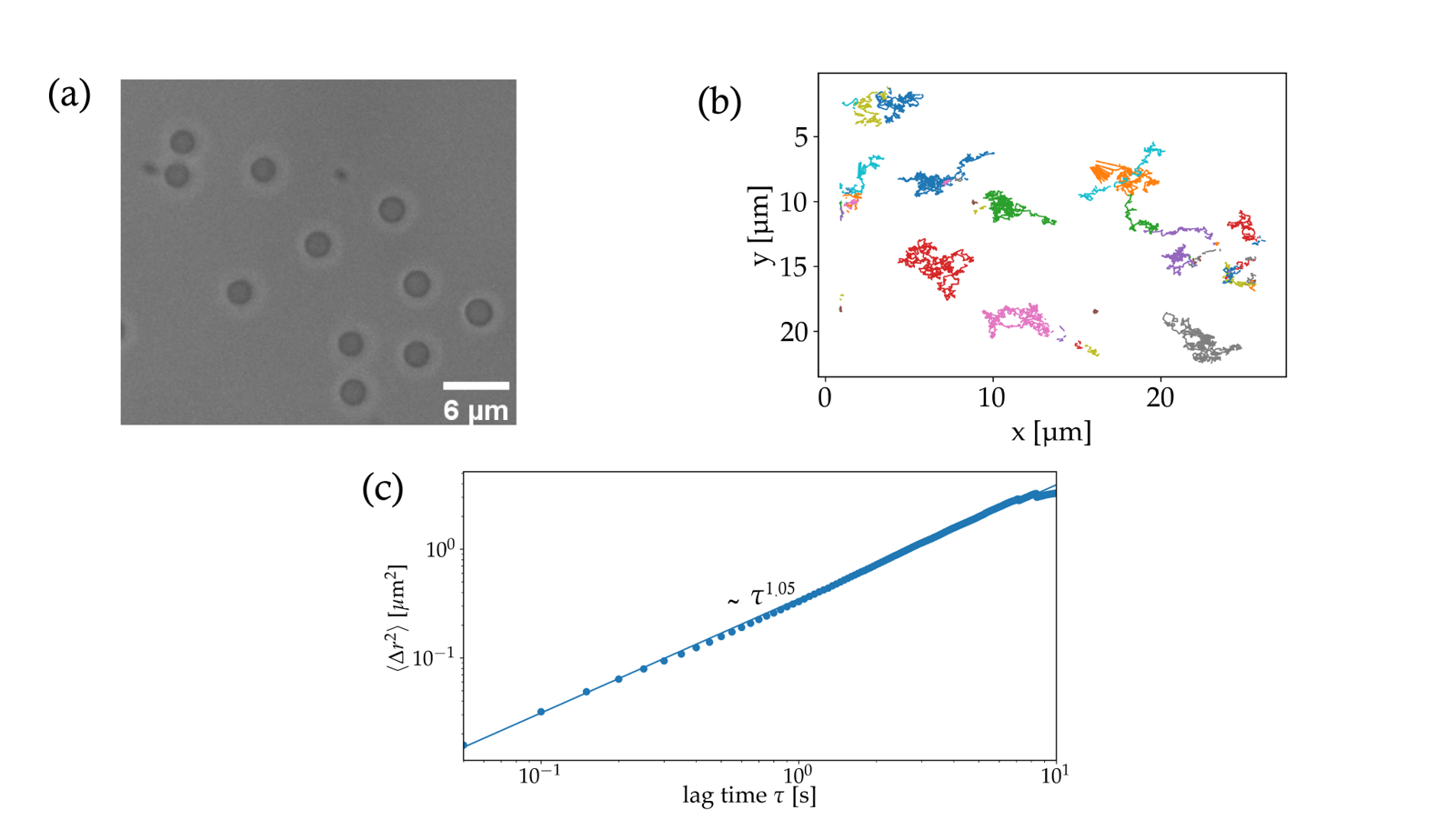}
  \caption{(a) Snapshot of particles undergoing Brownian motion. (b) Trajectories of the particles over a period of 50 s. (c) Ensemble MSD of all the particles. The diffusion exponent is 1.05. The Brownian motion of the particles is characterized when the particles are under thermal equilibrium i.e. there is no temperature gradient. The laser is switched off and the motion of the particles was recorded for 50 s. The trajectories of the particles is shown in figure S2 (b). MSD of all the particles is averaged and plotted against lag time on a log-log scale as shown in figure S2 (c). A diffusion coefficient of 1.05 shows that the particles are undergoing Brownian motion.}
  
\end{figure}

\section{S3: Single particle transport along the nanowire}

\begin{figure}[H]
\centering
  \includegraphics[width=\linewidth]{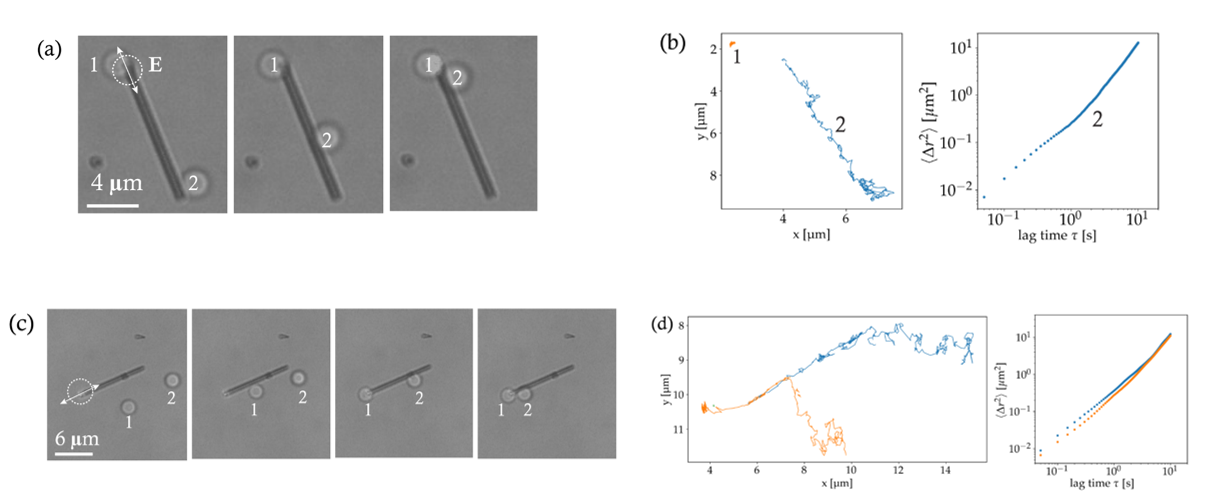}
  \caption{(a) Transport of a 2 $\mu$m silica colloid along the nanowire. The two particles are marked 1 and 2 for clear distinction between them. Particle 1 is already trapped at the excitation point. The corresponding trajectories of the two particles are plotted in S3 (b). As particle 1 is trapped at the excitation point, its trajectory is localized and is represented by orange color. The trajectory of particle 2 is shown by blue color. The corresponding MSD of the particle 2 is also shown. 
Taking another case, two particles reaching the nanowire at different locations is also considered. Figure S3 (c) shows the series of images of two particles. Particle 1 is trapped by the wire near its middle point after which it starts to move towards the excitation point, whereas particle 2 is captured at the distal end. The trajectories of both the particles are shown in (d) marked by orange (particle 1) and blue (particle 2) color. The corresponding MSD of both the particles when they are trapped by the wire is shown in the next panel of figure S3 (d). 
}
  
\end{figure}

\pagebreak

\section{S4: Near-field electric field distribution}

\begin{figure}[H]
\centering
  \includegraphics[width=\linewidth]{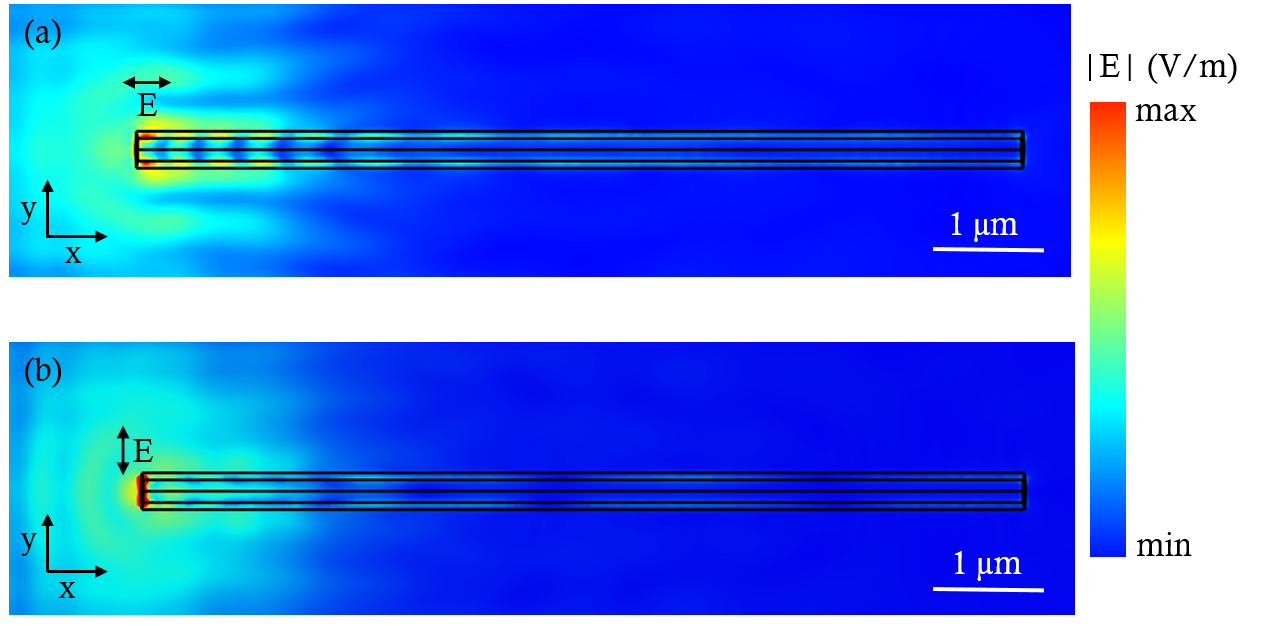}
  \caption{Calculated near-field electric field magnitude when one end of the nanowire is excited with polarization along the nanowire and (b) perpendicular to the nanowire.
}
  
\end{figure}

\pagebreak

\section{S4: Details on Supporting Movies}
\begin{enumerate}
  \item Supporting Movie 1: Transport of a single 2 $\mu$m silica colloid from the distal end of the nanowire to the excitation point
  
  \item Supporting Movie 2: 2 $\mu$m silica colloids undergoing Brownian motion.
  
  \item Supporting Movie 3: No trapping of 2.2 $\mu$m polystyrene beads is observed.
  
  \item Supporting Movie 4: Assembly process of 2 $\mu$m silica colloids.
  
  \item Supporting Movie 5: Assembly of silica beads when the polarization is along the nanowire.
  
  \item Supporting Movie 6: Assembly of silica beads when the polarization is perpendicular to the nanowire.

\end{enumerate}

\end{document}